\documentclass[12pt]{article}
\usepackage{amsmath,amssymb,color}
\usepackage{cite,epsf} 
\usepackage{pstricks}
\usepackage{color}
\usepackage{hyperref}
\usepackage{tikz}     

\usetikzlibrary{
decorations.pathmorphing %library needed for e.g. decoration=snake
}

\usepackage{graphicx,array} 
\newcommand{\be}{\begin{equation}}
\newcommand{\ee}{\end{equation}}
\newcommand{\beq}{\begin{equation}}
\newcommand{\eeq}{\end{equation}}
\newcommand{\bea}{\begin{eqnarray}}
\newcommand{\eea}{\end{eqnarray}}

\newcommand{\ba}{\begin{eqnarray}}
\newcommand{\ea}{\end{eqnarray}}

\def\sin{\mbox{sin}}
\def\cos{\mbox{cos}}

\def\log{\mbox{log}}

\begin{document}

\begin{titlepage}
\vspace{10pt}
\hfill
{\large\bf HU-EP-19/11}
\vspace{20mm}
\begin{center}

{\Large\bf   More on Wilson loops for two touching circles 
}

\vspace{45pt}

{\large Harald Dorn 
{\footnote{dorn@physik.hu-berlin.de
 }}}
\\[15mm]
{\it\ Institut f\"ur Physik und IRIS Adlershof, 
Humboldt-Universit\"at zu Berlin,}\\
{\it Zum Gro{\ss}en Windkanal 6, D-12489 Berlin, Germany}\\[4mm]

\vspace{20pt}

\end{center}
\vspace{10pt}
\vspace{40pt}

\centerline{{\bf{Abstract}}}
\vspace*{5mm}
\noindent
We calculate both at leading weak and strong coupling the renormalised
Maldacena-Wilson loop for  contours formed by consecutive passage of  two touching circles. At the touching point both circles 
should have the same normal direction but form cusps of non-zero
opening angle $\alpha$. Particular emphasis is put on the behaviour in the limit
$\alpha\rightarrow 0$ and its comparison with the spiky situation studied in a previous paper, where $\alpha$ was set to zero before renormalisation.

\vspace*{4mm}
\noindent

\vspace*{5mm}
\noindent
   
\end{titlepage}
\newpage

%\tableofcontents \newpage

\section{Introduction}
The study of ultraviolet divergences of Wilson loops due to cusps and self-intersections of their contours has a long history. It started
in the early eighties \cite{Polyakov:1980ca},\cite{Brandt:1981kf}, and over the years a lot of information about
the corresponding weak coupling perturbation theory has been accumulated. Of particular interest has been the cusp anomalous dimension, which is of relevance also in various other physical settings see e.g. \cite{Grozin:2015kna} and references therein.
The behaviour at strong coupling became accessible with the $AdS$-CFT holography \cite{Maldacena:1998im},\cite{Rey:1998ik},\cite{Drukker:1999zq} and is a  subject of ongoing interest.

The cusp anomalous dimension, and therefore also the renormalised Wilson loops for cusped contours diverge, if the
opening angle of the cusp tends to zero. Only recently we started the investigation of Wilson loops for contours, which
have a zero opening cusp, i.e. a spike, from the very beginning \cite{Dorn:2018als,Dorn:2018srz}. \footnote{The related problem for the holographic treatment of entanglement entropies has been discussed recently in \cite{Bueno:2019mex},\cite{Ghasemi:2019hdi}.}

In the first paper
we found a new type of ultraviolet divergence, which  is proportional to the inverse of the square root of the cutoff times the jump in the curvature at the tip of the spike.
The second paper is devoted to the renormalised  Wilson loops, i.e. the finite pieces remaining after subtraction of UV divergences and removal of the regularisation.
To have an example where analytical results can be obtained, the analysis is performed for two touching circles with opposite orientation at the touching point.
In this case, besides the new spike divergence, no logarithmic divergence appears. Usually the subtraction of logarithmic divergences  requires the introduction
of a RG-scale and thus a corresponding freedom for the renormalised quantities. But now no such RG-freedom appears in this case, and the 
renormalised Wilson loop turned out to be equal to one, both in lowest order weak as well as strong coupling.

To motivate the study of the present paper, let us make a small detour via the renormalisation issue for local composite operators. If such composite operators
are properly renormalised, their correlation functions are finite as long as the insertion points of the operators are all separated. If two such 
points approach each other, the corresponding correlation function develops a short distance singularity. In the coincidence limit a new composite operator
is formed. Its renormalisation can be treated either by starting with the coincidence case and renormalise afterwards or by using an operator product expansion
to control the short distance expansion. The renormalised version of the new composite is then defined by subtraction of the divergent short distance terms.
Up to the freedom in the choice of the RG scale both procedures yield the same result. 

Let us  now look at the case of a renormalised Wilson loop for a path formed by consecutive passage of  two touching circles. At the touching points both circles 
should have the same normal direction but form cusps of non-zero
opening angle $\alpha$. It will diverge for $\alpha\rightarrow 0$. This is the analogue to the previous paragraph. In both cases
there appear divergences in limits where a geometrical datum approaches zero, here the angle $\alpha$, there the distance between the insertion points
of the operators. The task of this paper is to check, whether via this route one gets the same result for the $\alpha =0$ case as in \cite{Dorn:2018srz}.
As a welcome byproduct of this analysis we will gain the renormalised Wilson loop for two touching circles at $\alpha\neq 0$. This extends the not so
large list of contours for which explicit analytical results are known. Even for the
special case of contours formed with circles the known results concern the coaxial case only, see \cite{Correa:2018pfn} and references therein. 

We will treat the local supersymmetric Maldacena-Wilson loop in ${\cal N}=4$ SYM theory 
\beq
W~=~\frac{1}{N}\big \langle \mbox{tr}~ P\mbox{exp} \int \big (iA_{\mu}\dot x^{\mu}+\vert \dot x\vert \phi_I\theta^I\big )d\tau \big \rangle~. \label{W-loop}
\eeq
The paper is organised as follows. Section 2 is devoted to
the lowest order weak coupling contribution to the renormalised Wilson loop for two touching circles of different radii $R_1>R_2$ and $\alpha\neq 0$. 
Besides the divergence for $\alpha\rightarrow 0$ it becomes also divergent for $R_1\rightarrow R_2$. This is the reason to start with equal radii before renormalisation in section 3. Furthermore, this section
contains the discussion of the $\alpha\rightarrow 0$ behaviour for both the unequal as well as the equal radii case. 
In section 4 we use the holographic formula \cite{Maldacena:1998im},\cite{Rey:1998ik} 
\beq
\mbox{log}~W~=~-\frac{\sqrt{\lambda}}{2\pi}~A~,\label{loop-malda}
\eeq
to relate the Wilson loop at strong 't Hooft coupling $\lambda=g^2N$ to the area $A$ of the minimal surface in $AdS$ approaching the Wilson loop contour at its boundary. Here we will succeed only in the equal radii case, where we can make use of a suitable conformal map to the straight line cusp  \cite{Drukker:1999zq}.
Section 5 is devoted to a summary and some conclusions. Several technical details of the calculations are presented in four appendices.
%%%%%%%%%%%%%%%%%%%%%%%%%%%%%%%%%%%%%%%%%%%%%
\section{Lowest order at weak coupling in ${\cal N}=4$ SYM}
%%%%%%%%%%%%%%%%%%%%%%%%%%%%%%%
We start with two circles
\bea
\vec x_1(\varphi_1)&=&R_1\big(\sin\varphi_1,~1-\cos\varphi_1,~0\big)~,\nonumber\\
\vec x_2(\varphi_2)&=&R_2\big (\cos\alpha~\sin\varphi_2,~(1-\cos\varphi_2),~\sin\alpha~\sin\varphi_2 \big )~,\label{x-circles}
\eea
with
\beq
0~<~\alpha~<~\pi~,~~~~~R_1>R_2~.\label{alpha}
\eeq
Then the contour to be  used in \eqref{W-loop}
is given by
\bea
\vec x(\tau)&=&\vec x_1(\tau)~,~~~~~~~~~~~~0\leq\tau\leq2\pi~,\nonumber\\
\vec x(\tau)&=&\vec x_2(4\pi-\tau)~,~~~~2\pi\leq\tau\leq 4\pi~.\label{contour}
\eea
For simplicity only the case with constant $\theta_I$ will be considered.

\begin{figure}[h!]
%\label{fig1}  
\begin{center}
  \includegraphics[width=8cm]{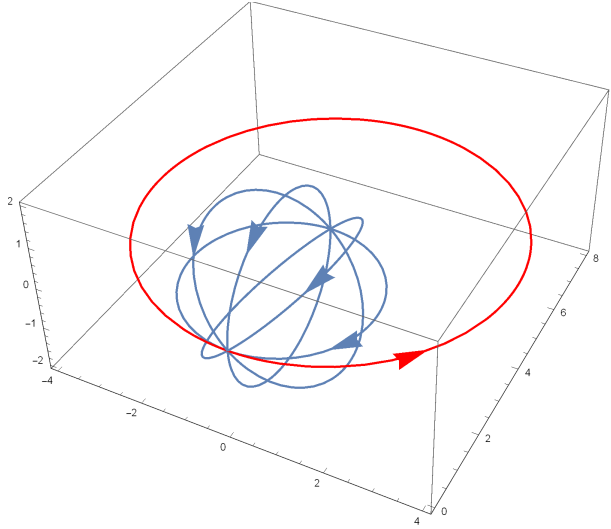}
  \end{center}

\caption {\it In red a larger circle with $R_1=4$. In blue  smaller circles with $R_2=2$ for various angles $\alpha =0,\frac{\pi}{4} ,\frac{\pi}{2},\frac{3\pi}{4}.$}
\label{fig1}
\end{figure}

For $\alpha=0$ one has the spiky situation of two touching circles with a  common tangent but opposite orientation as studied in \cite{Dorn:2018als,Dorn:2018srz }. For
$\alpha =\pi $ the circles have again a common tangent, but now with the same orientation. In the case $R_1=R_2$ the circles have two touching points.
For a visualisation see figure \ref{fig1}. The lowest order perturbative Wilson loop
has the structure \footnote{$C_F=\frac{N^2+1}{2N}$ for $SU(N)$ gauge group.}
\beq
\mbox{log}W~=~\frac{g^2C_F}{4\pi^2}\Big (I_{1}~+~I_{2}~+~I_{12}\Big )~+~{\cal O}(g^4)~.
\label{logW}
\eeq
%%%%%%%%%%%%%%%%%%%%%%%%%%%%%%%%%%%%%%%%%%%%
%%%%%%%%%%%%%%%%%%%%%%%%%%%%%%%%%%%
As in \cite{Dorn:2018als,Dorn:2018srz } we regularise the propagators by replacing
$(\vec x_1-\vec x_2)^2$ by $(\vec x_1-\vec x_2)^2+\epsilon^2$.
Then the trivial integrals $I_j$ are evaluated as
\beq
I_1~=~I_2~=~\pi^2~+~{\cal O}(\epsilon)~.\label{Ij}
\eeq
$I_{12}$ is given by 
\beq
I_{12}~=~\int_0^{2\pi}\int_0^{2\pi}~\frac{(1+\cos\alpha~\cos\varphi_1\cos\varphi_2+\sin\varphi_1\sin\varphi_2)~d\varphi_1~d\varphi_2}{
  D(\varphi_1,\varphi_2,\kappa,\alpha,\delta)}\label{I12original}~,
\eeq
with
\beq
D~=~2\big(\kappa(1-\cos\varphi_1)+\frac{1}{\kappa}(1-\cos\varphi_2)-\cos\alpha~\sin\varphi_1\sin\varphi_2-(1-\cos\varphi_1)(1-\cos\varphi_2)\big )+\delta^2
\eeq
and the dimensionless quantities \footnote{The limiting case of equal radii will be
  discussed in the next section.}
\beq
\delta~=~\frac{\epsilon}{\sqrt{R_1R_2}}~,~~~\kappa~=~\frac{R_1}{R_2}~>~1~.\label{kappa-del}
\eeq
Performing the $\varphi_2$-integration we get
\beq
I_{12}~=~I_{12}^{(1)}~+~I_{12}^{(2)}~,\label{I12split}
\eeq
with
\beq
I_{12}^{(1)}~=~\pi~\cos\alpha\int_0^{2\pi}\frac{(1-\frac{1}{\kappa})\cos\varphi -1}{(1-\frac{1}{\kappa})^2-2(1-\frac{1}{\kappa})\cos\varphi+\cos^2\alpha+\sin^2\alpha~\cos^2\varphi}~d\varphi~\label{I12-1}
\eeq
and
\beq
I_{12}^{(2)}~=~2\pi\int_0^2\frac{\Big (1+\cos\alpha~\frac{1+\frac{\kappa\delta^2}{2}+(A+\frac{\kappa\delta^2}{2}(\kappa-1))y+By^2}{1+Py+Qy^2}\Big )~~dy}{\sqrt{y(2-y)}\sqrt{\frac{\delta^4}{4}+\frac{\delta^2}{\kappa}+(2\sin^2\alpha+\delta^2(\kappa-1))y+((\kappa-1)^2-\sin^2\alpha)y^2}}~.\label{I12-2}
\eeq
In the last integral we have performed the change of variables
\beq
y~=~(1-\cos\varphi_1)
\eeq
and introduced the abbreviations
\bea
A&=&\kappa^2-1~,~~~~~~~~~~~~~~~~~~~~~
B~=~\kappa(\kappa-1)^2~,\nonumber\\
P&=&2\kappa^2(\cos^2\alpha-\frac{1}{\kappa})~,~~~~~~~~~~
Q~=~\kappa^2\sin^2\alpha ~.\label{A-Q}
\eea
The term $I_{12}^{(1)}$ is manifestly independent of $\delta$. It turns out to be
also independent of $\kappa$, equal to
\beq
I_{12}^{(1)}~=~2\pi^2\mbox{sign}\big (\alpha-\frac{\pi}{2}\big )~.\label{I12-1final}
\eeq
The second term  $I_{12}^{(2)}$ diverges if $\delta\rightarrow 0$, i.e. if the regularisation is removed.

To control its behaviour in this limit we write
\beq
I_{12}^{(2)}~=~I_{12}^{(2),\mbox{\scriptsize lead}}~+~I_{12}^{(2),\mbox{\scriptsize rest}}~,\label{I12-2split}
\eeq
with
\beq
I_{12}^{(2),\mbox{\scriptsize lead}}~=~2\pi\int_0^2\frac{(1+\cos\alpha)~~dy}{\sqrt{y(2-y)}\sqrt{\frac{\delta^2}{\kappa}+2y~\sin^2\alpha }}
\eeq
and $I_{12}^{(2),\mbox{\scriptsize rest}} $ defined by making in the integrand of $I_{12}^{(2)} $ the corresponding subtraction.

Now we obtain
\beq
I_{12}^{(2),\mbox{\scriptsize lead}}~=~\frac{\pi(1+\cos\alpha)}{\sin{\alpha}}~\log\frac{64 \kappa~\sin^2\alpha}{\delta^2}~+~{\cal O}(\delta^2\log\delta)~.\label{I12-2-leadfinal}
\eeq
In the limit $I_{12}^{(2),\mbox{\scriptsize rest}} $ stays finite and, to extract its
value, we can put $\delta = 0$ under the integral. This means
\beq
I_{12}^{(2),\mbox{\scriptsize rest}}~=~\frac{\sqrt{2}~\pi}{\sin\alpha}~\int_0^2\frac{dy}{y\sqrt{2-y}}\left ( \frac{1+\cos\alpha~\frac{1+Ay+By^2}{1+Py+Qy^2}}{\sqrt{1+S~ y}}-
(1+\cos\alpha)\right )~+~{\cal O}(\delta^2)~,\label{I12rest}
\eeq
with
\beq
S~=~\frac{(\kappa-1)^2-\sin^2\alpha}{2~\sin^2\alpha}~.\label{S}
\eeq
Some details of the evaluation of this integral one can find in appendix A. With \eqref{I12restfinal} from that appendix and \eqref{I12-2-leadfinal},\eqref{I12-2split},\eqref{I12-1final},\eqref{I12split} as well as \eqref{kappa-del}  we arrive at
\beq
I_{12}~=~-2\pi^2~+~4\pi\alpha~+~\frac{2\pi(1+\cos\alpha)}{\sin\alpha}~\log\Big (\frac{8~\sin^2\alpha}{\epsilon~\vert\frac{1}{R_1}-\frac{1}{R_2}\vert}\Big ) ~+~{\cal O}(\epsilon^2\log\epsilon)~.\label{I12final}
\eeq
Using this together with \eqref{logW},\eqref{Ij} and \eqref{I12final} we get
finally
\beq
\mbox{log}W~=~\frac{g^2C_F}{4\pi^2}\left (4\pi\alpha~+~\frac{2\pi(1+\cos\alpha)}{\sin\alpha}~\log\Big (\frac{8~\sin^2\alpha}{\epsilon~\vert\frac{1}{R_1}-\frac{1}{R_2}\vert} \Big )\right ) ~+~{\cal O}(\epsilon^2\log\epsilon)~.
\eeq
The logarithmic divergent term $\propto\log\epsilon$ can be obtained also by studying the case of two crossing straight lines with an orientation generating two touching cusps.

As usual in situations with a logarithmic divergence, there is a renormalisation group ambiguity for defining the renormalised quantity. With a RG-scale $\mu$ we subtract
the term $\propto\log(\epsilon\mu)$ and
get then  for the renormalised Wilson loop
\beq
\mbox{log}W_{\mbox{\scriptsize ren}}~=~\frac{g^2C_F}{4\pi^2}\left (4\pi\alpha~+~\frac{2\pi(1+\cos\alpha)}{\sin\alpha}~\log\Big (\frac{8\mu~\sin^2\alpha}{\vert\frac{1}{R_1}-\frac{1}{R_2}\vert} \Big )\right )~+~{\cal O}(g^4)~.\label{WrenR1R2}
\eeq
The discussion of the $\alpha$-dependence will be postponed to the end of the next
section, where it will be combined with that of the equal radii case.
%%%%%%%%%%%%%%%%%%%%%%%%%%%%%%%%%%%%%%%%%%
\section{The case of equal radii}
%%%%%%%%%%%%%%%%%%%%%%%%%%%%%%%%%%%
The case of equal radii $R:=R_1=R_2$, i.e. $\kappa=1$, requires separate treatment, since one is confronted with four cusps instead of two.\footnote{More precise: Two touching cusps and one self-intersection.} The integrals $I_1,~I_2$ and $ I_{12}^{(1)}$ are independent of the radii and can be taken from the previous
section. However, the integrand in \eqref{I12-2}, defining $I_{12}^{(2)}$, has now a non-integrable singularity not only at $y=0$, but also at $y=2$. Moreover, the integrand depends on $y(2-y)$ only, and we can write
\beq
I_{12}^{(2)}\big \vert_{\kappa=1}~=~4\pi~\int_0^1\frac{\big (1+\cos\alpha~\frac{1+\frac{\delta^2}{2}}{1-y(2-y)~\mbox{\scriptsize sin}^2\alpha}\big )~dy}{\sqrt{y(2-y)}\sqrt{\frac{\delta^4}{4}+\delta^2+y(2-y)\sin^2\alpha}}~.
\eeq
Now we split
\beq
I_{12}^{(2)}\big\vert_{\kappa=1}~=~\Big (I_{12}^{(2)}\big\vert_{\kappa=1}\Big )^{\mbox{\scriptsize lead}}~+~\Big (I_{12}^{(2)}\big\vert_{\kappa=1}\Big)^{\mbox{\scriptsize rest}}~,
\label{I12k-split}
\eeq
with
\beq
\Big (I_{12}^{(2)}\big\vert_{\kappa=1}\Big )^{\mbox{\scriptsize lead}}~=~4\pi\int_0^1\frac{(1+\cos\alpha)~~dy}{\sqrt{y(2-y)}\sqrt{\delta^2+2y~\sin^2\alpha }}
\eeq
and  $~\Big (I_{12}^{(2)}\big\vert_{\kappa=1}\Big)^{\mbox{\scriptsize rest}} $ defined by making in the integrand of $I_{12}^{(2)}\big\vert_{\kappa=1}$  the corresponding subtraction.  Then with manipulations similar to that in the previous section we get
\bea
\Big (I_{12}^{(2)}\big\vert_{\kappa=1}\Big )^{\mbox{\scriptsize lead}}&=& \frac{4\pi(1+\cos\alpha)}{\sin\alpha}~\log\Big (\frac{8 ~\sin\alpha}{\delta \sqrt{3+2\sqrt{2}}}\Big )~+~{\cal O}(\delta ^2\log\delta)~,\label{I12k-final}\\
\Big (I_{12}^{(2)}\big\vert_{\kappa=1}\Big )^{\mbox{\scriptsize rest}}&=&\frac{4\pi(1+\cos\alpha)}{\sin\alpha}~\log\Big (\frac{\sqrt{3+2\sqrt{2}}}{2}\Big )+4\pi\alpha-4\pi^2~\Theta(\alpha-\frac{\pi}{2})+{\cal O}(\delta^2)~.\nonumber
\eea
Collecting now \eqref{logW},\eqref{Ij},\eqref{kappa-del},\eqref{I12split},\eqref{I12-1final},\eqref{I12k-split}, and \eqref{I12k-final}, the final result is
\beq
\mbox{log}W\big\vert_{\kappa=1}~=~\frac{g^2C_F}{4\pi^2}\left (4\pi\alpha~+~\frac{4\pi(1+\cos\alpha)}{\sin\alpha}~\log\Big (\frac{4R~\sin\alpha}{\epsilon} \Big )\right ) ~+~{\cal O}(\epsilon^2\log\epsilon)~
\eeq
and
\beq
\mbox{log}\big (W\big\vert_{\kappa=1}\big )_{\mbox{\scriptsize ren}}~=~\frac{g^2C_F}{4\pi^2}\left (4\pi\alpha~+~\frac{4\pi(1+\cos\alpha)}{\sin\alpha}~\log(4R\mu~\sin\alpha)\right ) ~+~{\cal O}(g^4)~.\label{WrenR}
\eeq
\\

Comparing now this result with \eqref{WrenR1R2} we find
\beq
\mbox{log}W_{\mbox{\scriptsize ren}}(\alpha)~=~\mbox{log}\big (W(\alpha)\vert_{\kappa=1}\big )_{\mbox{\scriptsize ren}}~~~~~~~\mbox{if}~~~~~~\frac{1}{(\mu R)^2}~=~2\left \vert\frac{1}{\mu R_1}-\frac{1}{\mu R_2}\right \vert~.
\eeq
Therefore we comment on the $\alpha$-dependence of
\beq
w(\alpha,\nu)~=~ \frac{1}{\pi}~\Big (\alpha ~+~\frac{1+\cos\alpha}{\sin\alpha}~\log(4\nu~\sin \alpha)\Big )~.
\eeq
Up to the factor $g^2C_F$ this yields \eqref{WrenR} if $\nu=\mu R$ and \eqref{WrenR1R2}  if $\nu=\sqrt{\frac{\mu R_1}{2(\kappa-1)}}$.

\begin{figure}[h!]
%\label{fig1a}  
\begin{center}
\includegraphics[width=11cm]{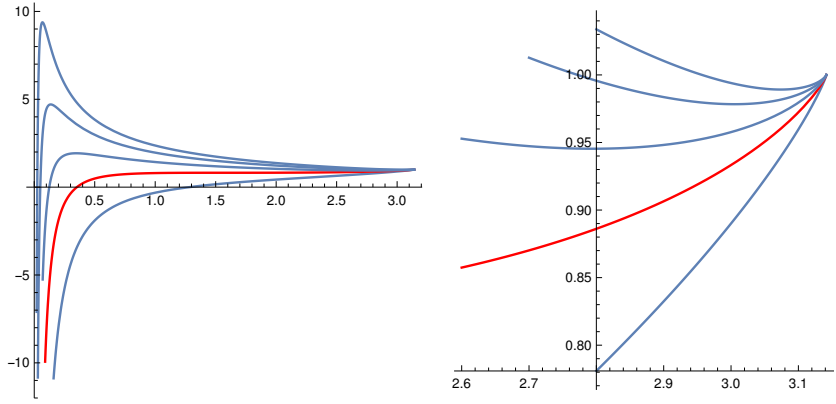}
\end{center}

\caption{\it The function $w(\alpha,\nu)$ for $\nu= 10,5,2,\frac{\displaystyle e}{4},0.1$ (from above), on the left in the whole $\alpha$-interval, on the right zoomed into the vicinity of $\alpha =\pi$. The case $\nu=\frac{\displaystyle e}{4}$ is shown in red.  }
\label{fig1a}
\end{figure}

Now for {\it all} $\nu$
\beq
w(\pi,\nu)~=~1~,~~~~~w(0,\nu)~=~-\infty ~.
\eeq
$w(\alpha,\nu)$ is a monotonic rising function of $\alpha\in(0,\pi)$ as long as $\nu<\frac{\displaystyle e}{4}$. If $\nu>\frac{\displaystyle e}{4}$ there is a maximum \footnote{Just above $\frac{\displaystyle e}{4}$
  only a local maximum, but soon the absolute maximum in $\alpha\in(0,\pi)$.}
at $\alpha=\mbox{arcsin}\frac{\displaystyle e}{4\nu}$ and a local minimum at $\alpha=\pi-\mbox{arcsin}\frac{\displaystyle e}{4\nu}$. For some illustration see figure \ref{fig1a}.
%%%%%%%%%%%%%%%%%%%%%%%%%%%%%%%%%%%%%%%%%%%
%%%%%%%%%%%%%%%%%%%%%%%%%%%%%%%%%%%%%%%

%%%%%%%%%%%%%%%%%%%%%%%%%%%%%%%%%%%%%%%%%%%%%
\section{Holographic evaluation at strong coupling}
\begin{figure}[h!]
%\label{fig2}  
\begin{center}
\includegraphics[width=9cm]{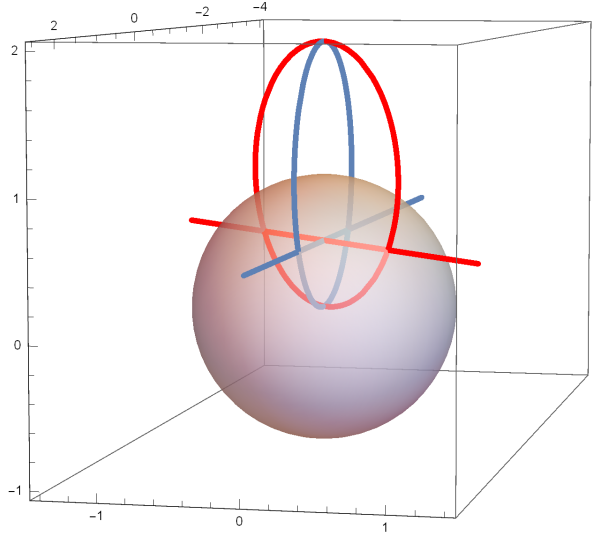}
\\
\end{center}
\caption{\it A transparent unit sphere with parts of the straight  lines  and their circular images. $R=1$ and $\alpha=\pi/4$.  }
\label{fig2}
\end{figure}

%%%%%%%%%%%%%%%%%%%%%%%%%%%%%%%
For generic radii $R_1\neq R_2$, we do not know of any explicit construction of the minimal surface in $AdS$, relevant for the holographic evaluation of our Wilson loop. However, in the special case of equal radii one can generate the two circles as the image under inversion on the unit sphere of two  straight lines crossing each other at an angle $\alpha$. The intersection point has to have a distance $1/(2R)$ from the centre of the unit sphere, and the lines should be both orthogonal to the straight line connecting the intersection point with the centre.  For a visualisation see figure \ref{fig2}.
%%%%%%%%%%%%%%%%%%%%%%%%%%%%%%%%%%%%%%%%%%%
%%%%%%%%%%%%%%%%%%%%%%%%%%%%%%%%%%%%%%%
The minimal surface approaching at the boundary of $AdS$ the cusp formed by the two half-lines, with positive $x_1$ and $x_2$, is in Poincare coordinates  given by \cite{Drukker:1999zq},
\bea
x_1&=&\rho~\cos\varphi~,~~~~~~
x_2~=~\rho ~\sin\varphi~,~~~~~~x_3~=~\frac{1}{2R}~,\nonumber\\[2mm]
z&=&\frac{\rho}{f(\varphi)}~,~~~~~~~0\leq\varphi\leq\alpha~,~~~0\leq\rho<\infty~.\label{cusp}
\eea
The function $f(\varphi)$ has the property
\beq
f(\alpha-\varphi)~=~f(\varphi)~\label{phi-f-sym}
\eeq
and is for $0\leq\varphi\leq\frac{\alpha}{2}$ defined  implicitly by
\beq
\varphi~=~f_0\sqrt{1+f_0^2}~\int_{f(\varphi)}^{\infty}\frac{df}{\sqrt{(f^4+f^2)^2-(f_0^4+f_0^2)(f^4+f^2)}}~.\label{phi-f}
\eeq
Finally, the parameter$f_0$ is related to the opening angle $\alpha$ by
\beq
f_0~=~f\big (\frac{\alpha}{2}\big )~.\label{alpha-f0}
\eeq
Extending now the inversion on the unit sphere to an isometry inside $AdS$, we find the part of the minimal surface related
to the  two half-circles with positive $x_1,x_2$. \footnote{The other part is obtained by $x_j\rightarrow -x_j, ~j=1,2$ in \eqref{surface}. Both parts are separated up to the two touching points of the circles.\label{twoparts}}
\bea
x_1&=&\frac{4R^2\rho~\cos\varphi}{4R^2\rho^2(1+f^{-2})+1}~,~~~~~x_2~=~\frac{4R^2\rho~\sin\varphi}{4R^2\rho^2(1+f^{-2})+1}~,\nonumber\\
x_3&=&\frac{2R}{4R^2\rho^2(1+f^{-2})+1}~,~~~~~~~z~=~\frac{4R^2\rho}{f(4R^2\rho^2(1+f^{-2})+1)}~.\label{surface}
\eea

\begin{figure}[h!]
%\label{fig3}  
\begin{center}
\includegraphics[width=14cm]{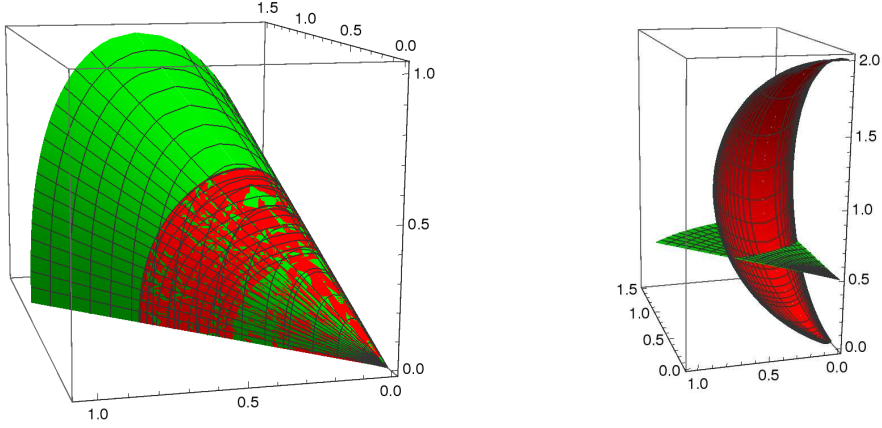}
\end{center}

\caption{\it Projections on the $(x_1,x_2,z)$-subspace (left) and on the $(x_1,x_2,x_3)$-subspace (right), in red for the surface under discussion, in green
  for its preimage, green extends to infinity, for red the whole surface is shown, $R=1$ and $\alpha=\frac{\pi}{4}$.}
\label{fig3}
\end{figure}

For a visualisation of this surface in $AdS_4$ we have to rely on projections onto three-dimensional subspaces. Before presenting
corresponding figures, it is useful to take notice of the following  facts. 
\bea
x_j (\hat\rho,\varphi)&=&x_j(\rho,\varphi )~,~~~~j=1,2~,\nonumber\\
x_3 (\hat\rho,\varphi)&=&2R~-~x_3 (\rho,\varphi)~,\nonumber\\
z (\hat\rho,\varphi )&=&z (\rho,\varphi )~,\label{rho-phi-sym}
\eea
with
\beq
\hat\rho~=~\frac{f(\varphi)^2}{4\rho~R^2(1+f^2)}~.
\eeq
Furthermore, the surface parameter point $(\rho,\varphi)$ on \eqref{surface} has the same $x_1,x_2$ and\\[2mm]
 $z$-coordinate
as the parameter point $\big (\frac{4R^2f^2\rho}{f^2+4\rho^2R^2(1+f^2)},\varphi\big )$ on the preimage \eqref{cusp}.\\[2mm]
These analytic properties help to understand the projections obtained numerically for the case $R=1$ and $\alpha =\pi/4$, as shown in figure \ref{fig3}.
%%%%%%%%%%%%%%%%%%%%%%%%%%%
%%%%%%%%%%%%%%%%%%%%%%%%%%%%%%%%%%%%%%%%%

\begin{figure}[h!]
%\label{fig4}  
\begin{center}
\includegraphics[width=9cm]{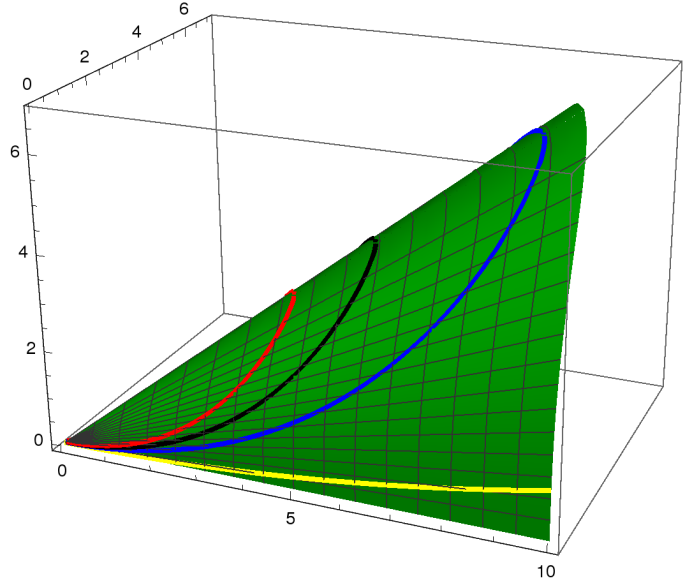}
\end{center}

\caption{\it Preimages of the boundary of the cutted surface \eqref{surface}  for \newline $\epsilon=0.1,0.075,0.05,0.01$ in red, black, blue, yellow. Again 
$R=1,~\alpha=\frac{\pi}{4}$.}
\label{fig4}
\end{figure}

The regularised area $A_{\epsilon} $, needed for the holographic evaluation of our Wilson loop, is
now given by the double of the area of the surface \eqref{surface} cutted at $z=\epsilon$.  The factor two takes into account the second part, mentioned in footnote \ref{twoparts}. The boundary curve of the cutted surface is in surface coordinates $(\rho,\varphi)$ defined by
\beq
\epsilon~=~~\frac{4R^2\rho f(\varphi)}{4R^2\rho^2(1+f^{2})+f^2}~.\label{cut}
\eeq
The induced metric looks simpler on the preimage \eqref{cusp}. Therefore we prefer
to take advantage of the isometry property of the map between \eqref{cusp} and \eqref{surface} and calculate on the preimage. The preimage of the boundary curve \eqref{cut} for some values of $\epsilon$ is shown in figure \ref{fig4}.
%%%%%%%%%%%%%%%%%%%%%%%%%%%%%%%%%%%%%%%%%%%
%%%%%%%%%%%%%%%%%%%%%%%%%%%%%%%%%%

Taking the induced metric on the preimage \eqref{cusp} from \cite{Drukker:1999zq} we
get (${\cal B}_{\epsilon}$ denoting the range of $(\rho,\varphi)$ for which the
r.h.s. of \eqref{cut} is larger than $\epsilon$.)
\beq
A_{\epsilon}~=~2~\int_{{\cal B}_{\epsilon}}\frac{\sqrt{f^4+f^2+(f')^2}}{\rho}~d\rho~ d\varphi~.
\eeq
We now change the integration variable $\varphi$ to $f$, taking into account
the symmetry \eqref{phi-f-sym} and arrive at
\bea
A_{\epsilon}&=&4~\int_{f_0}^{f_{\epsilon}}df\int_{\rho^-_{\epsilon}}^{\rho^+_{\epsilon}}\frac{d\rho}{\rho}\sqrt{\frac{f^4+f^2}{f^4+f^2-f_0^4-f_0^2}}\nonumber\\[2mm]
&=&4~\int_{f_0}^{f_{\epsilon}}U(f,f_0)~\log~\frac{\rho^+_{\epsilon}(f)}{\rho^-_{\epsilon}(f)}~df~,
\eea
with
\beq
\rho^{\pm}_{\epsilon}(f)~=~\frac{f}{2\epsilon (1+f^2)}~\left(1\pm\sqrt{1-\frac{\epsilon^2}{R^2}(1+f^2)}\right )~,\label{rhopm}
\eeq
\beq
f_{\epsilon}~=~\sqrt{\frac{R^2}{\epsilon^2}-1}~,\label{feps}
\eeq
\beq
U(f,f_0)=\sqrt{\frac{f^4+f^2}{f^4+f^2-f_0^4-f_0^2}}\label{Sff0}~.
\eeq
Using $\rho^{+}_{\epsilon}\rho^{-}_{\epsilon}=\frac{f^2}{4R^2(1+f^2)}$ and again the notation $\delta=\epsilon/R$ this lead to
\beq
A_{\epsilon}~=~A^{(1)}_{\epsilon}~+~A^{(2)}_{\epsilon}~+~A^{(3)}_{\epsilon}~,\label{Aeps-split}    
\eeq
where the $A_{\epsilon}^{(j)}$ are given by
\bea
A^{(1)}_{\epsilon}&=&8~\int_{f_0}^{\sqrt{\frac{1}{\delta^2}-1}} U(f,f_0)~\log\big(1+\sqrt{1-\delta^2(1+f^2)}\big )~df~,\label{Aeps1-def}\\
A^{(2)}_{\epsilon}&=&-~8~\log\delta ~\int_{f_0}^{\sqrt{\frac{1}{\delta^2}-1}} U(f,f_0)~df~,\\
~A^{(3)}_{\epsilon}&=&-~4~\int_{f_0}^{\sqrt{\frac{1}{\delta^2}-1}} U(f,f_0)~\log(1+f^2)~df~.
\eea
Straightforward estimates yield
\bea
A^{(2)}_{\epsilon}&=&4~ \Gamma_{\mbox{\scriptsize cusp}}~\log\delta~-~8~\frac{\log\delta}{\delta}~+~{\cal O}(\delta\log\delta)~,\label{Aeps2}\\
A^{(3)}_{\epsilon}&=&8~\frac{\log\delta}{\delta}~+~\frac{8}{\delta}~+~4f_0\log(1+f_0^2)+8~\mbox{arctan}f_0-4\pi-8f_0 \nonumber\\[2mm]
&&~~-~4~\int_{f_0}^{\infty}\big  (U(f,f_0)-1\big )~\log(1+f^2)~df~+~{\cal O}(\delta\log\delta)~. \label{Aeps3}
\eea
Above we introduced the strong coupling cusp anomalous dimension \cite{Drukker:1999zq}, see also \cite{Dorn:2015bfa},
\bea
\Gamma_{\mbox{\scriptsize cusp}}(\alpha)&=&2f_0~-~2\int_{f_0}^{\infty}\big (U(f,f_0)-1\big )~df~\label{Gammacusp}\\
&=&\frac{\pi}{2}\frac{f_0^2}{\sqrt{1+f_0^2}}~_2F_1\Big(\frac{1}{2},\frac{3}{2},2,\frac{-f_0^2}{1+f_0^2}\Big )~.\nonumber
\eea
A little bit more effort is necessary for $A^{(1)}_{\epsilon}$. It is discussed in appendix B with the result
\beq
A^{(1)}_{\epsilon}~=~\frac{4\pi-8}{\delta}~-~4~\log2~\Gamma_{\mbox{\scriptsize cusp}}~+~{\cal O}(\sqrt{\delta})~.\label{Aeps1}
\eeq
Now inserting \eqref{Aeps2},\eqref{Aeps3},\eqref{Aeps1} into \eqref{Aeps-split} we  arrive at
\beq
A_{\epsilon}~=~\frac{4\pi R}{\epsilon} ~-~4\Gamma_{\mbox{\scriptsize cusp}}~\log\frac{2R}{\epsilon}~+A_0~+~{\cal O}(\sqrt{\epsilon})~,\label{Aeps-final}
\eeq
with
\beq
A_0~=~4f_0\log(1+f_0^2)+8~\mbox{arctan}f_0-4\pi-8f_0-4\int_{f_0}^{\infty}\big (U(f,f_0)-1\big )\log(1+f^2)df.\label{A0}
\eeq
$\Gamma_{\mbox{\scriptsize cusp}}$ and $A_0$ are via \eqref{phi-f} and \eqref{alpha-f0} functions of the cusp angle $\alpha$. 

\begin{figure}[h!]
%\label{fig5}  
%\begin{centering}
\includegraphics[width=14cm]{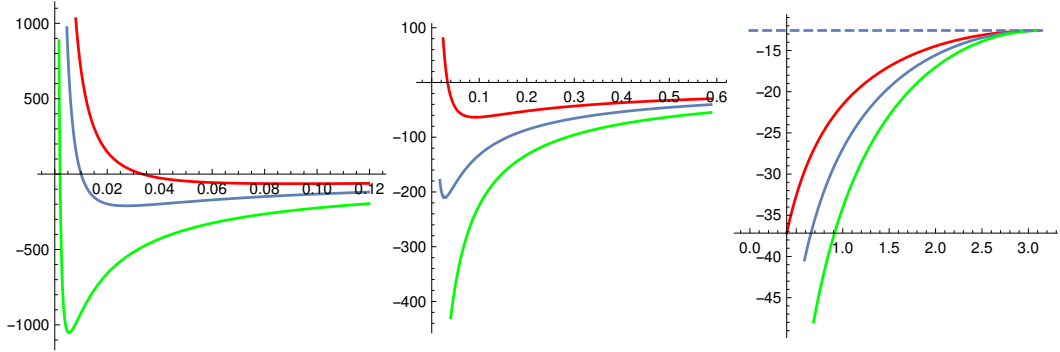}
\\
%\end{centering}
\caption{\it $A_{\mbox{\scriptsize ren}}$ as a function of $\alpha$ in different pieces of the interval $0\leq\alpha\leq\pi$ and for $\mu R=3,10,50$ in red, blue, green. The plots are generated by ParametricPlot with $f_0$ as parameter. The dashed line
  in the right picture indicates the value $-4\pi$.  }
\label{fig5}
\end{figure}

For the construction of a renormalised area, we again have
 to handle the ambiguity in the subtraction of a logarithmic divergent term
by introducing a RG-scale $\mu$
\beq
A_{\mbox{\scriptsize ren}}~=~A_0~-~4\Gamma_{\mbox{\scriptsize cusp}}~\log(2R\mu)~.\label{Aren}
\eeq
A numeric evaluation for three different values of $\mu R$ can be seen in figure \ref{fig5}.
%%%%%%%%%%%%%%%%%%%%%%%%%%%%%%%%%%%%%%%%%%%

For the discussion below the behaviour of $A_{\mbox{\scriptsize ren}}$ near $\alpha =0$
and $\alpha =\pi$ is of special interest.
With the results of appendix C we get using \eqref{alpha-pi},\eqref{Gamma-pi} and \eqref{Mpi}
\beq
A_{\mbox{\scriptsize ren}}~=~-4\pi~+~{\cal O}(\pi-\alpha)~.\label{Aren-pi}
\eeq
Near $\alpha=0$ we find with \eqref{alpha-zero},\eqref{Gamma-zero},\eqref{a},\eqref{A0-alpha-zero}
\bea
A_{\mbox{\scriptsize ren}}&=&-4a b~\frac{\log\alpha}{\alpha}~-~\Big (b(8+a_2)+4a b~\log\frac{2R\mu}{b}\Big )~\frac{1}{\alpha}~+~{\cal O}(\alpha\log\alpha)~\label{Arenalpha-null}\\
&=&-5.742'~\frac{\log\alpha}{\alpha}~-~\Big (9.214' ~+~5.742' ~\log(2R\mu)\Big )~\frac{1}{\alpha}~+~{\cal O}(\alpha\log\alpha)~.\nonumber
\eea
The strong oscillation of $A_{\mbox{\scriptsize ren}}$ in figure \ref{fig5}  near $\alpha=0$ is
due to the fact that the leading divergent term is positive while the nextleading divergent term is negative. \footnote{ As long as $R\mu > \frac{b}{2}\exp(-\frac{8+a_2}{4a})=0.10048\dots$ .}\\

In \cite{Dorn:2015bfa} the  renormalised area of the minimal surface for a boundary
contour in a plane, which is  composed out of two circular arcs of radii $R_1,R_2$
forming cusps of angle $\alpha$ with a distance $D$ between the tips of the cusps,
has been calculated. Comparing it with our present result for a contour built from
two semi-circles with equal radius $R$ and forming cusps with the same opening
angle (i.e. one half of \eqref{Aren}) one finds complete agreement
since in our case $D=2R$.

Both contours can be generated as an image under inversion
on the unit sphere of a suitable placed single cusp between two straight half-lines
with angle $\alpha$. Therefore, the two just compared contours can be conformally
mapped to each other. Moreover, both contours are special cases of a whole
set of contours with two cusps of opening angle $\alpha$ composed out of
circular arcs. Due to unbroken Poincar$\acute{ \mbox{e}}$ invariance the
corresponding renormalised Wilson loops will be a function of $\mu,D,R_1,R_2,\alpha$. The only conformally invariant parameter in this set is $\alpha$. The breaking
of conformal invariance is due to the presence of the cusps. Since the cusp
anomalous dimension depends also in the case of curved wings on $\alpha$ only \cite{Dorn:2015bfa},
one should expect that the symmetry breaking term in $A_{\mbox{\scriptsize ren}}$ or
$W_{\mbox{\scriptsize ren}}$, respectively, depends only on $\mu D$ and $\alpha$.
As a whole one  would have a conformal covariant expression, i.e. an invariant
form for a function of $\mu D$ and $\alpha$ which changes its value under
conformal maps only via changes of $D$. A partial check of this conjecture
is given by the comparison of the two special cases above.

The situation resembles that for lightlike polygons. There is a symmetry
breaking term controlled by the anomalous conformal Ward identities and, as soon
as conformal invariant parameters are available, in addition a conformal invariant remainder
function \cite{Drummond:2007au}. In the tetragon case there is even another aspect of analogy.  All tetragons can be generated by a conformal map of the lightlike
straight half-line cusp \cite{Kruczenski:2002fb},\cite{Alday:2007hr}.

%%%%%%%%%%%%%%%%%%%%%%%%%%%%%%%%%%%%%%%
\section{Summary and discussion}
%%%%%%%%%%%%%%%%%%%%%%%%%%%%%%%
Concerning the motivating question posed in the introduction, our main result
is the observation, that in the limit $\alpha\rightarrow 0$ in all three cases \eqref{WrenR1R2},\eqref{WrenR} and \eqref{Arenalpha-null} (via \eqref{loop-malda}) $\log W_{\mbox{\scriptsize ren}}$ beyond the diverging terms has {\it no} nonzero finite contribution. Hence the procedure  \{$\alpha\neq 0$, renormalise, expand for $\alpha\rightarrow 0$, subtract divergent terms, $\alpha =0$\}  yields the same result as the procedure \{$\alpha =0$, renormalise\} as used in \cite{Dorn:2018srz}.    

Another common feature of all three cases is found for $\alpha\rightarrow\pi $. While
$\log W_{\mbox{\scriptsize ren}}$ depends for $0<\alpha<\pi$ on the RG-scale and the radii, it becomes independent of these parameters at $\alpha =\pi$. In the latter case
the cusps disappear, and one has a smooth contour at hand, the doubly wounded circle. In the weak coupling cases
of sections 2 and 3 one gets then four times the result for a single circle. The renormalised  minimal area \eqref{Aren-pi} becomes twice that for a single circle.\footnote{This observation on the weak and strong coupling limits (note the square root in \eqref{loop-malda}) is consistent with the all order result $W_{\mbox{\scriptsize double}}(\lambda)=W_{\mbox{\scriptsize single}}(4\lambda)$, see \cite{Drukker:2000rr, Pestun:2007rz}.}

For large $\mu R$ both the weak and strong coupling results in sections 3 and 4
show a remarkable strong oscillation near $\alpha =0$, due to the opposite sign of the leading and nextleading term.

This opposite sign in the strong coupling case holds
for all $\mu R$, but in the weak coupling case only for large enough $\mu R$. 
This is a clear indication, that the interpolation for $\log W_{\mbox{\scriptsize ren}}$ between weak and strong coupling requires a genuine function of $g^2,~\alpha$ and $\mu R$,
which {\it cannot} be factorised in a product $F(g^2)H(\alpha,\mu R)$.  One finds in appendix D some more detailed  discussion of this issue.

Further work related to the issues raised in this paper should concern the study of higher order corrections, both at weak and strong coupling. It would also be very interesting to elaborate the anomalous conformal Ward identities as acting on general polygon like contours whose edges are circular arcs. As indicted at the end of the last section this could deliver important structural information on Wilson loops for this subset of contours.
\\[15mm]
{\bf Acknowledgement:}\\[5mm]
I would like to thank the 
Quantum Field and String Theory Group at Humboldt University
for kind hospitality.\\[15mm]
%%%%%%%%%%%%%%%%%%%%%%%%%%%%%%%%

%%%%%%%%%%%%%%%%%%%%%%%%%%%%%%%%
\section*{Appendix A}
The indefinite integral related to \eqref{I12rest} is
\bea
J(\kappa,\alpha,y)~=~\frac{2~\pi}{\sin\alpha}\left  ( 2 ~\cos^2\frac{\alpha}{2}\Big ( \mbox{arctanh}\sqrt{\frac{2}{2-y}}-\mbox{arctanh}\sqrt{\frac{2(1+Sy)}{2-y}}\Big )\right .\nonumber\\
 ~~~~~~~~~~~~~~~~\left . +~
f^+~\mbox{arctan}\Big (g^+\sqrt{\frac{1+Sy}{2-y}}\Big )~-~f^-~\mbox{arctan}\Big (g^-\sqrt{\frac{1+Sy}{2-y}}\Big )\right )~,\label{J}
\eea
with
\bea
f^{\pm}~=~\cos\alpha ~\cdot~~~~~~~~~~~~~~~~~~~~~~~~~~~~~~~~~~~~~~~~~~~~~~~~~~~~~~~~~~~~~~~~~~~~~~~~~~~~~~~~~~~~~~~~\\[2mm]
\frac{P^2 +2 B (P+1) +2Q(P-1)-A(P+4Q) \pm (2B+A-P-2Q)\sqrt{P^2-4Q} }{\sqrt{
 P^2 - 4 Q}\sqrt{1 + 2 P + 4 Q} \sqrt{
 2 S -4Q + (2 S-1) P \pm(2S+1) \sqrt{P^2 - 4 Q} }  }~,\nonumber
\eea
\beq
g^{\pm}~=~\frac{\sqrt{2+4P+8Q}}{\sqrt{2S-4Q+P(2S-1)\pm(2S+1)\sqrt{P^2-4Q}}}~.
\eeq
These long expressions simplify tremendously after insertion of \eqref{A-Q} and \eqref{S}
\bea
f^+&=&\mbox{sign}(\cos\alpha)~\sin\alpha~,\label{fp}\\
f^-&=&f^+~\mbox{sign}\Big (\frac{(\kappa-1)^2}{\kappa^2+(\kappa-1)^2}-\sin^2\alpha\Big )~,\label{fm}
\eea
\beq
g^+=\frac{\sqrt{2}~(2\kappa-1)~\sin\alpha}{\kappa\sqrt{(\kappa-1)^2-\kappa^2\sin^2\alpha}+(\kappa-1)^2\vert\cos\alpha\vert}~,\label{gp}
\eeq
\beq
g^-=\frac{\sqrt{2}~(2\kappa-1)~\sin\alpha}{\kappa\sqrt{(\kappa-1)^2-\kappa^2\sin^2\alpha}-(\kappa-1)^2\vert\cos\alpha\vert}~
\mbox{sign}\Big ( \frac{(\kappa-1)^2}{\kappa^2+(\kappa-1)^2}-\sin^2\alpha \Big )~.\label{gm}
\eeq
Since the arctan-function is odd, the explicit sign-factors in \eqref{fm} and \eqref{gm} cancel, and we can write the second line of \eqref{J} as (and understanding below $g^-$ without the sign-factor in \eqref{gm}) 
$$f^+~\left (\mbox{arctan}\Big (g^+\sqrt{\frac{1+Sy}{2-y}}\Big )~-~\mbox{arctan}\Big (g^-\sqrt{\frac{1+Sy}{2-y}}\Big )\right )~.$$
Then with an addition theorem for the arctan-function we get
\bea
J(\kappa,\alpha,y)~=~\frac{2~\pi}{\sin\alpha}\left  ( 2 ~\cos^2\frac{\alpha}{2}\Big ( \mbox{arctanh}\sqrt{\frac{2}{2-y}}-\mbox{arctanh}\sqrt{\frac{2(1+Sy)}{2-y}}\Big )\right .\nonumber\\
 ~~~\left . +~f^+~n~\pi ~+~
f^+~\mbox{arctan}\Big (\frac{(g^+-g^-)\sqrt{\frac{1+Sy}{2-y}}}{1+\frac{1+Sy}{2-y}~ g^+g^-}\Big )\right )~.\label{Jmod}
\eea
Here $n$ is an integer, whose value can depend on $\alpha,~\kappa$ and $y$. In our integration interval the argument of the arctanh-functions
in \eqref{Jmod} is larger than 1. Since one is free to add an arbitrary constant in the indefinite integral we can replace in \eqref{Jmod}
$\mbox{arctanh}(x)=\frac{1}{2}\log\frac{1+x}{1-x}$ by $ \frac{1}{2}\log\frac{x+1}{x-1}$.

With
\beq
\frac{\sqrt{2}~(g^+-g^-)}{2+ g^+g^-}~=~-~\mbox{sign}(\cos\alpha)~\mbox{arctan}\big (\mbox{tan}(2\alpha)\big )
\eeq
we then get for \eqref{I12rest}, i.e. $J(\kappa,\alpha,2)-J(\kappa,\alpha,0)$,
\beq
I_{12}^{(2),\mbox{\scriptsize rest}}~=~4\pi~\Big (\alpha~-~\pi~\Theta\big (\alpha-\frac{\pi}{2}\big )~-~
\frac{1+\cos\alpha}{2~\sin\alpha}~\log\frac{\kappa-1}
{\sin\alpha}~\Big )~.\label{I12restfinal}
\eeq
The integer, left open so far in the discussion  above, has been determined by comparison with the numerical evaluation of \eqref{I12rest}. Its presence
results in the term with the unitstep-function.
%%%%%%%%%%%%%%%%%%%%%%%%%%%%%%%
\section*{Appendix B}
This appendix is devoted to the $\epsilon\rightarrow 0$ expansion of $A^{(1)}_{\epsilon}$ defined in \eqref{Aeps1-def}. To start with, we write it as (remember $\delta=\epsilon/R$ )
\bea
A^{(1)}_{\epsilon}&=&A^{(1,1)}_{\epsilon}~+A^{(1,2)}_{\epsilon}~,\label{A1-split}\\[2mm]
A^{(1,1)}_{\epsilon}&=&8\int_{f_0}^{\frac{\sqrt{1-\delta^2}}{\delta}}\big (U(f,f_0)-1\big )~\log\big (1+\sqrt{1-\delta^2(1+f^2)}\big )~df~,\\
    A^{(1,2)}_{\epsilon}&=&8\int_{f_0}^{\frac{\sqrt{1-\delta^2}}{\delta}}\log\big (1+\sqrt{1-\delta^2(1+f^2)}\big )~df.
\eea
The indefinite integral for $A^{(1,2)}_{\epsilon}$ is a certain combination of linear, logarithmic and arctan terms. Inserting the boundaries, a straightforward expansion yields
\beq
A^{(1,2)}_{\epsilon}~=~4~\frac{\pi-2}{\delta}~-~8f_0~\log 2~+~{\cal O}(\delta)~.\label{A12}
\eeq
$A^{(1,1)}_{\epsilon}$  can be written as
\bea
A^{(1,1)}_{\epsilon}&=&8~\log2~\int_{f_0}^{\frac{\sqrt{1-\delta^2}}{\delta}}\big (U(f,f_0)-1\big )~df\nonumber\\
&+&8\int_{f_0}^{\frac{\sqrt{1-\delta^2}}{\delta}}\big (U(f,f_0)-1\big )~\log\frac{ 1+\sqrt{1-\delta^2(1+f^2)}}{2}~df.
\eea
The first term in the last equation tends to the corresponding integral extended up to infinity plus a term ${\cal O}(\delta)$.
In the second term the small $\delta$ expansion of the log-term cannot be used uniformly in the whole integration region.
Therefore, we split it in two parts, one integral over the interval $(f_0,1/\sqrt{\delta})$ and one integral over the remainder.
Then in the first part the expansion of the log-term can be used, giving a contribution ${\cal O}(\delta)$. For the second
term we use the boundedness of the log-term in the whole integration region to establish an estimate ${\cal O}(\sqrt{\delta})$. Hence
\beq
A^{(1,1)}_{\epsilon}~=~8~\log 2\int_{f_0}^{\infty}\big  (U(f,f_0)-1\big )~df~+~{\cal O}(\sqrt{\delta})~.
\eeq
With \eqref{A1-split},\eqref{A12} and the definition of $\Gamma_{\mbox{\scriptsize cusp}}$ in \eqref{Gammacusp} we get
equation \eqref{Aeps1} in the main text.
%%%%%%%%%%%%%%%%%%%%%%%%%%%%%%%%%%%
\section*{Appendix C}
Here we analyse the dependence of $\Gamma_{\mbox{\scriptsize cusp}}$ and $A_0$ (see \eqref{Gammacusp} and \eqref{A0}) on the cusp angle $\alpha$. From  \eqref{phi-f} and \eqref{alpha-f0} we get
\bea
\alpha &=& \pi~+~{\cal O}(f_0)~~~~~~~~~~~~~~~~~~~~~~~~~~~~~~~\mbox{at}~~~f_0\rightarrow 0~,
\label{alpha-pi}\\[2mm]
\alpha &=& \frac{b}{f_0}~+~{\cal O}(f_0^{-3}))~, ~~~~b=\frac{(2\pi)^{\frac{3}{2}}}{(\Gamma(\frac{1}{4}))^2}~~~~~ \mbox{at}~~~f_0\rightarrow \infty ~.\label{alpha-zero}
\eea
Therefore, to control the behaviour at the boundaries of the $\alpha$ interval $(0,\pi)$ we have to look at the behaviour at $f_0\rightarrow \infty$ and  $f_0\rightarrow 0$, respectively.

With the substitution $f^2=f_0^2+z^2$ one can bring $\Gamma_{\mbox{\scriptsize cusp}}$
into the form \cite{Drukker:1999zq}
\beq
\Gamma_{\mbox{\scriptsize cusp}}~=~2\int_0^{\infty}\left( 1-\sqrt{\frac{z^2+1+f_0^2}{z^2+1+2f_0^2}}\right )dz~.
\eeq
From there one gets easily
\beq
\Gamma_{\mbox{\scriptsize cusp}}~=~\frac{\pi}{2}~f_0^2~+~{\cal O}(f_0^4)\label{Gamma-pi}
\eeq
and
\bea
\Gamma_{\mbox{\scriptsize cusp}}&=&a~f_0~+~{\cal O}(f_0^{-1})~,\label{Gamma-zero}\\[2mm]
a&=&2\int _0^{\infty}\Big (1-\sqrt{\frac{1+x^2}{2+x^2}}\Big )dx~=~2 E(-1)-\frac{(\Gamma (\frac{1}{4}))^2}{2\sqrt{2\pi}}~=~1.198\dots ~.\label{a} 
\eea
Next we study the asymptotics of the integral in \eqref{A0} (with the factor 4 included). After the same substitution as above it becomes
\beq
M(f_0)~=~4\int_0^{\infty}\left (\sqrt{\frac{z^2+1+f_0^2}{z^2+1+2f_0^2}}-\sqrt{\frac{z^2}{z^2+f_0^2}}\right )~\log(1+f_0^2+z^2)~dz~.
\eeq
This yields straightforwardly for $f_0\rightarrow 0$
\beq
M(f_0)~=~{\cal O}(f_0^2)~. \label{Mpi}
\eeq
For the other limit we write $M$ as
\beq
M(f_0)=4f_0\int_0^{\infty}\left (\sqrt{\frac{x^2+1+f_0^{-2}}{x^2+2+f_0^{-2}}}-\sqrt{\frac{x^2}{x^2+1}}\right )\log\big (f_0^2(1+x^2)+1\big )~dx
\eeq
and get
\beq
M(f_0)~=~a_1 f_0\log f_0~+~a_2 f_0~+~{\cal O}\Big (\frac{\log f_0}{f_0}\Big )~,\label{M-inf}
\eeq
with
\bea
a_1&=&8\int_0^{\infty}\left (\sqrt{\frac{x^2+1}{x^2+2}}-\sqrt{\frac{x^2}{x^2+1}}\right )dx~=~3.204\dots ~,\\
a_2&=&4\int_0^{\infty}\left (\sqrt{\frac{x^2+1}{x^2+2}}-\sqrt{\frac{x^2}{x^2+1}}\right )\log(1+x^2)dx~=~0.556\dots ~.
\eea
Both constants can be also expressed in terms of standard special functions
\bea
a_1&=&8~\Big (1-E(-1)+\frac{(\Gamma(\frac{1}{4}))^2}{4\sqrt{2\pi}}\Big )~=~8-4a~,\label{a1}\\
a_2&=&\frac{(4+\pi)(2\pi)^{\frac{3}{2}}}{(\Gamma(\frac{1}{4}))^2}~-~8~.
\eea

Inserting  \eqref{Mpi} and \eqref{M-inf},\eqref{a1} respectively  into \eqref{A0} we get for $f_0\rightarrow 0$
\beq
A_0~=~-4\pi~+~{\cal O}(f_0^2) \label{A0-f0-zero}
\eeq
and for $f_0\rightarrow\infty$
\beq
A_0~=~
4a~f_0\log f_0~-~(8+a_2)f_0~+~{\cal O}\Big (\frac{\log f_0}{f_0}\Big)~.\label{A0-alpha-zero}
\eeq
%%%%%%%%%%%%%%%%%%%%%%%%%%%%%%%%%%%
\section*{Appendix D}
\begin{figure}[h!]
%\label{fig7}  
\begin{center}
\includegraphics[width=12cm]{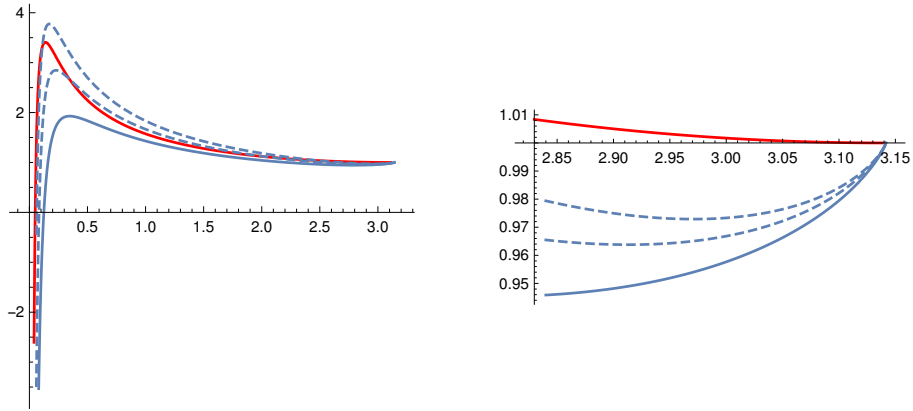}
\\
\end{center}
\caption{\it Comparison of the shape of the $\alpha$-dependence for strong coupling (red) and weak coupling (blue). The red and the solid blue curve are for $\mu R=2$. The  blue dashed curves are for $\mu R =3$ and $4$, respectively. The right picture is a zoom into the vicinity of $\alpha =\pi$.  }
\label{fig7}
\end{figure}

Here we add some comments comparing the shape of the $\alpha$-dependence of $\log W_{\mbox{\scriptsize ren}}$ in the equal radii case for weak and strong coupling.
If there would be for all couplings a factorisation $\log W_{\mbox{\scriptsize ren}}=F(g^2)H(\mu R,\alpha)$, then
$$\frac{\log W_{\mbox{\scriptsize ren}}(g^2,\mu R,\alpha)}{\log W_{\mbox{\scriptsize ren}}(g^2,\mu R,\pi)}$$
would be an universal shape function independent of the coupling and normalised to one at
$\alpha =\pi$.
With \eqref{WrenR}  we get at weak coupling
\beq
\left .\frac{\log W_{\mbox{\scriptsize ren}}(g^2,\mu R,\alpha)}{\log W_{\mbox{\scriptsize ren}}(g^2,\mu R,\pi)}\right \vert_{\mbox{\scriptsize weak}}~=~\frac{1}{\pi}~\Big (\alpha ~+~\frac{1+\cos\alpha}{\sin\alpha}~\log(4\mu R~\sin \alpha)\Big )~.
\eeq
At strong coupling holds via \eqref{loop-malda},\eqref{A0},\eqref{Aren},\eqref{Aren-pi}
\beq
\left .\frac{\log W_{\mbox{\scriptsize ren}}(g^2,\mu R,\alpha)}{\log W_{\mbox{\scriptsize ren}}(g^2,\mu R,\pi)}\right\vert_{\mbox{\scriptsize strong}}~=~-\frac{A_{\mbox{\scriptsize ren}}(\mu R,\alpha)}{4\pi}~.
\eeq
In figure \ref{fig7} we show a numeric plot of both the weak and strong coupling shape functions for $\mu R=2$. They differ clearly. Obviously also by playing with different RG-scales $\mu$ no agreement can be obtained. This is another check, that
there cannot exist an overall factorisation as asked for at the beginning of this appendix.
%%%%%%%%%%%%%%%%%%%%%%%%%%%%%%%%%%%%%%%%%%%%
%%%%%%%%%%%%%%%%%%%%%%%%%%%%%%%%

\begin{figure}[h!]
%\label{fig8}  
\begin{center}
\includegraphics[width=4cm]{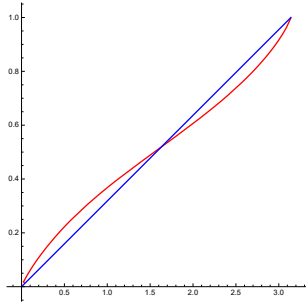}
\\
\end{center}
\caption{\it Comparison of the shape of the $\alpha$-dependence of candidates for a remainder function for strong coupling (red) and weak coupling (blue).}
\label{fig8}
\end{figure}

What could be possible candidates for a remainder function (and normalised to one at $\alpha =\pi$) in the sense of the discussion at the end of section 4 ?  
Looking at \eqref{WrenR} it could be $\frac{\alpha}{\pi}$ at weak coupling, and with \eqref{Aren},\eqref{A0},\eqref{A0-alpha-zero} 
$\frac{-A_0(f_0)+4a f_0\mbox{\scriptsize log} f_0-(8+a_2)f_0}{4\pi}$ at strong coupling. We plot both functions in figure \ref{fig8}. A zoom into
the vicinity of the crossing of both curves  shows that it is located near 1.63,  i.e. not at $\frac{\pi}{2}$.
%%%%%%%%%%%%%%%%%%%%%%%%%%%%%%%%%%%%%%%%%%%%
%%%%%%%%%%%%%%%%%%%%%%%%%%%%%%%%

%%%%%%%%%%%%%%%%%%%%%%%%%%%%%%%%


\newpage
\begin{thebibliography}{99}

\bibitem{Polyakov:1980ca}
  A.~M.~Polyakov,
  %``Gauge Fields as Rings of Glue,''
  Nucl.\ Phys.\ B {\bf 164} (1980) 171.
  %doi:10.1016/0550-3213(80)90507-6
  %%CITATION = doi:10.1016/0550-3213(80)90507-6;%%

\bibitem{Brandt:1981kf}
  R.~A.~Brandt, F.~Neri and M.~A.~Sato,
  %``Renormalization of Loop Functions for All Loops,''
  Phys.\ Rev.\ D {\bf 24} (1981) 879.
  %doi:10.1103/PhysRevD.24.879
  %%CITATION = doi:10.1103/PhysRevD.24.879;%%

\bibitem{Grozin:2015kna}
  A.~Grozin, J.~M.~Henn, G.~P.~Korchemsky and P.~Marquard,
  %``The three-loop cusp anomalous dimension in QCD and its supersymmetric extensions,''
  JHEP {\bf 1601} (2016) 140
 % doi:10.1007/JHEP01(2016)140
  [arXiv:1510.07803 [hep-ph]].
  %%CITATION = doi:10.1007/JHEP01(2016)140;%%

\bibitem{Maldacena:1998im}
  J.~M.~Maldacena,
  %``Wilson loops in large N field theories,''
  Phys.\ Rev.\ Lett.\  {\bf 80} (1998) 4859
 % doi:10.1103/PhysRevLett.80.4859
  [hep-th/9803002].
  %%CITATION = doi:10.1103/PhysRevLett.80.4859;%%

\bibitem{Rey:1998ik}
  S.~J.~Rey and J.~T.~Yee,
  %``Macroscopic strings as heavy quarks in large N gauge theory and anti-de Sitter supergravity,''
  Eur.\ Phys.\ J.\ C {\bf 22} (2001) 379
 % doi:10.1007/s100520100799
  [hep-th/9803001].
  %%CITATION = doi:10.1007/s100520100799;%%

\bibitem{Drukker:1999zq}
  N.~Drukker, D.~J.~Gross and H.~Ooguri,
  %``Wilson loops and minimal surfaces,''
  Phys.\ Rev.\ D {\bf 60} (1999) 125006
 % doi:10.1103/PhysRevD.60.125006
  [hep-th/9904191].
  %%CITATION = doi:10.1103/PhysRevD.60.125006;%%

 %\cite{Dorn:2018als} 
\bibitem{Dorn:2018als}
  H.~Dorn,
  %``On a new type of divergence for spiky Wilson loops and related entanglement entropies,''
  JHEP {\bf 1803} (2018) 124
   Erratum: [JHEP {\bf 1805} (2018) 108]
  %doi:10.1007/JHEP03(2018)124, 10.1007/JHEP05(2018)108
  [arXiv:1801.10367 [hep-th]].
  %%CITATION = doi:10.1007/JHEP03(2018)124, 10.1007/JHEP05(2018)108;%%

%\cite{Dorn:2018srz}
\bibitem{Dorn:2018srz}
  H.~Dorn,
  %``On Wilson loops for two touching circles with opposite orientation,''
  J.\ Phys.\ A {\bf 52} (2019) no.9,  095401
  %doi:10.1088/1751-8121/ab0003
  [arXiv:1811.00799 [hep-th]].
  %%CITATION = doi:10.1088/1751-8121/ab0003;%%

  %\cite{Bueno:2019mex}
\bibitem{Bueno:2019mex}
  P.~Bueno, H.~Casini and W.~Witczak-Krempa,
  ``Generalizing the entanglement entropy of singular regions in conformal field theories,''
  arXiv:1904.11495 [hep-th].
  %%CITATION = ARXIV:1904.11495;%%

%\cite{Ghasemi:2019hdi}
\bibitem{Ghasemi:2019hdi}
  M.~Ghasemi and S.~Parvizi,
  ``Curved Corner Contribution to the Entanglement Entropy in an Anisotropic Spacetime,''
  arXiv:1905.01675 [hep-th].
  %%CITATION = ARXIV:1905.01675;%%  

\bibitem{Correa:2018pfn}
  D.~Correa, P.~Pisani, A.~Rios Fukelman and K.~Zarembo,
  %``Dyson equations for correlators of Wilson loops,''
  JHEP {\bf 1812} (2018) 100
 % doi:10.1007/JHEP12(2018)100
  [arXiv:1811.03552 [hep-th]].
  %%CITATION = doi:10.1007/JHEP12(2018)100;%%
  

\bibitem{Dorn:2015bfa}
  H.~Dorn,
  %``Wilson loops at strong coupling for curved contours with cusps,''
  J.\ Phys.\ A {\bf 49} (2016) no.14,  145402
  %doi:10.1088/1751-8113/49/14/145402
  [arXiv:1509.00222 [hep-th]].
  %%CITATION = doi:10.1088/1751-8113/49/14/145402;%%  

\bibitem{Drummond:2007au}
  J.~M.~Drummond, J.~Henn, G.~P.~Korchemsky and E.~Sokatchev,
  %``Conformal Ward identities for Wilson loops and a test of the duality with gluon amplitudes,''
  Nucl.\ Phys.\ B {\bf 826} (2010) 337
 % doi:10.1016/j.nuclphysb.2009.10.013
  [arXiv:0712.1223 [hep-th]].
  %%CITATION = doi:10.1016/j.nuclphysb.2009.10.013;%%

\bibitem{Kruczenski:2002fb}
  M.~Kruczenski,
  %``A Note on twist two operators in N=4 SYM and Wilson loops in Minkowski signature,''
  JHEP {\bf 0212} (2002) 024
 % doi:10.1088/1126-6708/2002/12/024
  [hep-th/0210115].
  %%CITATION = doi:10.1088/1126-6708/2002/12/024;%%

\bibitem{Alday:2007hr}
  L.~F.~Alday and J.~M.~Maldacena,
  %``Gluon scattering amplitudes at strong coupling,''
  JHEP {\bf 0706} (2007) 064
 % doi:10.1088/1126-6708/2007/06/064
  [arXiv:0705.0303 [hep-th]].
  %%CITATION = doi:10.1088/1126-6708/2007/06/064;%%


\bibitem{Drukker:2000rr}
  N.~Drukker and D.~J.~Gross,
  %``An Exact prediction of N=4 SUSYM theory for string theory,''
  J.\ Math.\ Phys.\  {\bf 42} (2001) 2896
 % doi:10.1063/1.1372177
  [hep-th/0010274].
  %%CITATION = doi:10.1063/1.1372177;%%

\bibitem{Pestun:2007rz}
  V.~Pestun,
  %``Localization of gauge theory on a four-sphere and supersymmetric Wilson loops,''
  Commun.\ Math.\ Phys.\  {\bf 313} (2012) 71
 % doi:10.1007/s00220-012-1485-0
  [arXiv:0712.2824 [hep-th]].
  %%CITATION = doi:10.1007/s00220-012-1485-0;%%

\end{thebibliography}
\end{document}